# Order and Chaos:
# Collective Behavior of Crowded Drops in Microfluidic Systems

*Sindy Tang – Stanford University*

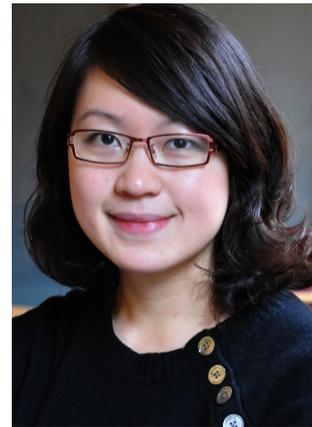

## Biography

*Dr. Sindy KY Tang joined the faculty of Stanford University in September 2011 as an assistant professor in the Department of Mechanical Engineering. She received her Ph.D. from Harvard University in Engineering Sciences under the supervision of Prof. George Whitesides. Her lab at Stanford works on the fundamental understanding of fluid mechanics and mass transport in microfluidic systems, and the application of this knowledge towards problems in biology, rapid diagnostics for health and environmental sustainability. The current areas of focus include the hydrodynamics of concentrated emulsions in confinements, interfacial mass transport and self-assembly, and ultrahigh throughput opto-microfluidic systems for biochemical sensing and diagnostics, water and energy sustainability, and single-cell wound healing studies. Dr. Tang's work has been recognized by multiple awards including the NSF CAREER Award, 3M Nontenured Faculty Award, and the ACS Petroleum Fund New Investigator Award. Web: http://web.stanford.edu/group/tanglab/*

## Abstract

Droplet microfluidics, in which micro-droplets serve as individual reactors, has enabled a range of high-throughput biochemical processes.[1] The talk will start with our recent application on using droplets to identify bacteria, specifically methane-metabolizing bacteria, for increasing the efficiency of generation of bioplastics.[2] This work, along with emerging applications to screen very large libraries of molecules or cells, has motivated us to investigate the physics and design criteria necessary for the further scaling-up of droplets technology.

Unlike solid wells typically used in current biochemical assays, droplets are subject to instability and can break especially at fast flow conditions. Although the physics of single drops has been studied extensively, the flow of crowded drops or concentrated emulsions—where droplet volume fraction exceeds ~80%—is relatively unexplored in microfluidics. The ability to leverage concentrated emulsions is critical for increasing the throughput of droplet applications, as it avoids the need to process large volumes of the continuous phase in order to process the same number of drops. Prior work on concentrated emulsions focused on their bulk rheological properties. The behavior of individual drops within the emulsion is not well understood, but is important as each droplet carries a different reaction.

This talk examines the collective behavior of drops in a concentrated emulsion by tracking the dynamics and the fate of individual drops within the emulsion. At the fast flow limit, we show that droplet breakup within the emulsion is stochastic (Figure 1).[3-6] This contrasts the deterministic breakup in classical single-drop studies. We further demonstrate that the breakup probability is described by dimensionless numbers including the capillary number and confinement factor, and the stochasticity originates from the time-varying packing configuration of the drops.[6]

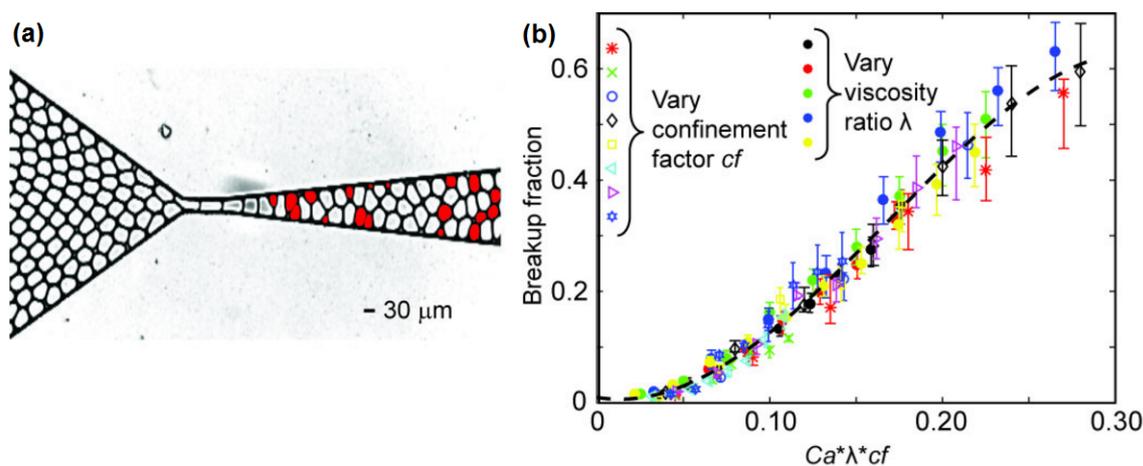

Figure 1. (a) Microscopy image of concentrated emulsion flowing in a tapered microchannel. Drops that underwent breakup after the constriction are highlighted in red. (b) Breakup fraction as a function of the product of capillary number (Ca), viscosity ratio (l) and confinement factor (cf).

To mitigate breakup, we design novel amphiphilic nanoparticles "F-SiO$_2$ NPs" (Figure 2a), and show they are more effective than surfactant molecules in preventing droplet breakup.[5] In addition, we show that these particles mitigate cross-talk of droplet content arising from inter-drop molecular transport, which is found to be mediated by reverse micelles in droplet systems stabilized by surfactants (Figure 2b).[7-11] The particles also form a solid-like interface that supports the adhesion and growth of adherent cells (Figure 2c). This capability is not possible in surfactant systems, and opens new opportunities for the use of droplets to study an increased range of cell types that require rigid surfaces for adhesion.

At the slow flow limit of the concentrated emulsion, we observe an unexpected order, where the velocity of individual drops in the emulsion exhibits spatiotemporal periodicity.[12] Such periodicity is surprising from both fluid and solid mechanics point of view. We show the phenomenon can be explained by treating the emulsion as a soft crystal undergoing plasticity, in a nanoscale system comprising thousands of atoms as modeled by droplets. Our results represent a new type of collective order not described before, and have practical use in on-chip droplet manipulation. From the solid mechanics perspective, the phenomenon directly contrasts the stochasticity of dislocations in microscopic crystals, and suggests a new approach to control the mechanical forming of nanocrystals.

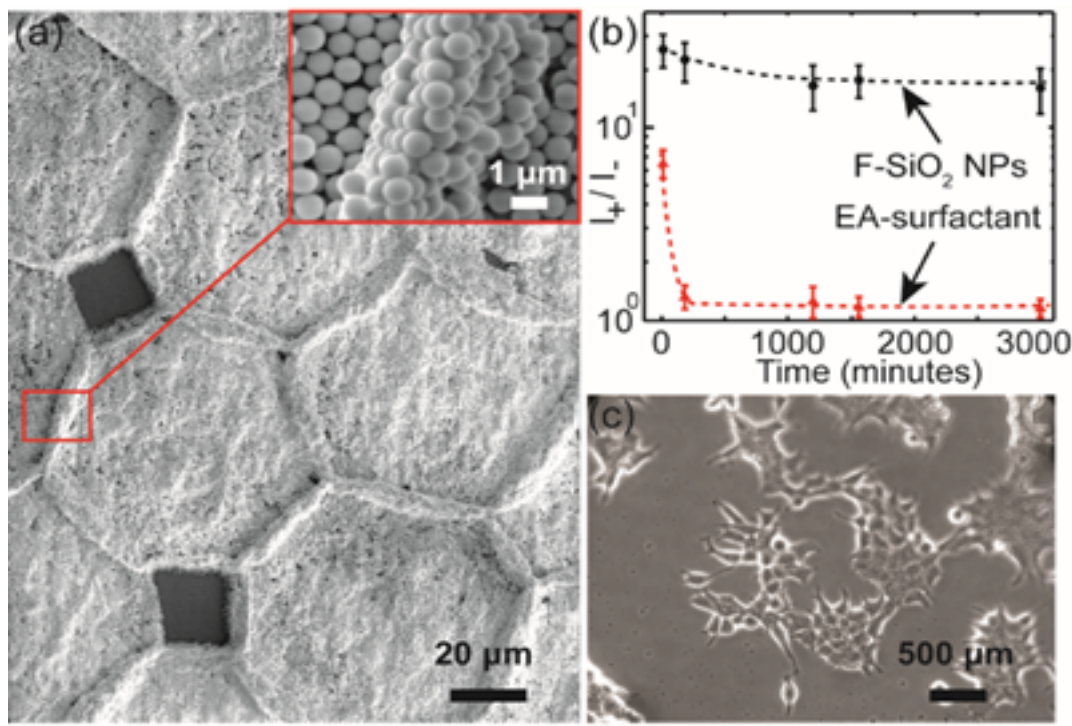

*Figure 2. (a) SEM image of F-SiO2 NPs stabilized droplets after evaporation of fluids. (b) Plots showing the time evolution of resorufin's fluorescence intensity ratio between positive droplets (contains 220 µM resorufin at t=0) and negative droplets (without resorufin at t=0). (c) Optical image showing the spread and growth of anchorage-dependent mammalian cells (MCF-7) after 19 hours of cultivation at F-SiO2 NPs stabilized water/HFE-7500 interface.*

|  | High stability? | Mitigate cross-talk? | Proliferation of adherent cells? | Functionalizable? |
|---|---|---|---|---|
| **F-SiO$_2$ NPs** *Patent pending* | Yes* | Yes | Yes | Yes |
| **Surfactant** | Yes* | No | No | Difficult |

*Figure 3. Comparison of nanoparticles and surfactant as droplet stabilizer.*

# References

1. Liat Rosenfeld, Tiras Lin, Ratmir Derda, and Sindy K.Y. Tang, "Review and Analysis of Performance Metrics of Droplet Microfluidics Systems," *Microfluidics and Nanofluidics*, 16, 5, 921-939, 2014.

2. Jaewook Myung, Minkyu Kim, Ming Pan, Craig S. Criddle, and Sindy K.Y. Tang, "Low energy emulsion-based fermentation enabling accelerated methane mass transfer and growth of poly(3-hydroxybutyrate)-accumulating methanotrophs", *Bioresource Technology*, 207, 302–307, 2016.

3. Liat Rosenfeld, Lin Fan, Yunhan Chen, Ryan Swoboda, and Sindy K.Y. Tang, "Break-up of Droplets in a Concentrated Emulsion Flowing through a Narrow Constriction," *Soft Matter*, 10, 421-430, 2014.

4. Ya Gai, Jian Wei Khor, and Sindy K. Y. Tang, "Confinement and viscosity ratio effect on droplet break-up in a concentrated emulsion flowing through a narrow constriction", *Lab on a Chip*, 16, 3058-3064, 2016.

5. Ya Gai, Minkyu Kim, Ming Pan, and Sindy K. Y. Tang, "Amphiphilic nanoparticles suppress droplet break-up in a concentrated emulsion flowing through a narrow constriction", *Biomicrofluidics*, 11, 034117, 2017.

6. Jian Wei Khor, Minkyu Kim, Simon Schütz, Tobias Schneider, and Sindy K.Y. Tang, "Time-varying droplet configuration determines break-up probability of drops within a concentrated emulsion", *Applied Physics Letters*, 111, 124102, 2017.

7. Yunhan Chen, Adi W. Gani, and Sindy K.Y. Tang, "Characterization of Sensitivity and Specificity in Leaky Droplet-based Assays," *Lab on a Chip*, 12, 5093 – 5103, 2012.

8. Ming Pan, Liat Rosenfeld, Minkyu Kim, Manqi Xu, Edith Lin, Ratmir Derda, and Sindy K.Y. Tang, "Fluorinated Pickering Emulsions Impede Interfacial Transport and Form Rigid Interface for the Growth of Anchorage-dependent Cells", *ACS Applied Materials & Interfaces*, 6, 21446-21453, 2014.

9. Ming Pan, Fengjiao Lyu, and Sindy K.Y. Tang, "Fluorinated Pickering Emulsions with Non-adsorbing Interfaces for Droplet-based Enzymatic Assays", *Analytical Chemistry*, 87, 7938–7943, 2015.

10. Ming Pan, Minkyu Kim, Luke Blauch, and Sindy K.Y. Tang, "Surface-Functionalizable Amphiphilic Nanoparticles for Pickering Emulsions with Designer Fluid-Fluid Interfaces", *RSC Advances*, 6, 39926-39932, 2016.

11. Ming Pan, Fengjiao Lyu, and Sindy K. Y. Tang, "Methods to Coalesce Fluorinated Pickering Emulsions", *Analytical Methods*, DOI: 10.1039/C7AY01289F, 2017.

12. Ya Gai, Chia Leong, Wei Cai, and Sindy K. Y. Tang, "Spatiotemporal periodicity of dislocation dynamics in a two-dimensional microfluidic crystal flowing in a tapered channel", *Proceedings of the National Academy of Sciences*, 113, 12082-12087, 2016.